\documentclass[12pt,preprint]{aastex}

\newcommand\rfe{\hbox{$r/{\rm Fe}$}}
\newcommand\feh{\hbox{${\rm Fe}/{\rm H}$}}
\newcommand\rpro{{\em r}-process}
\newcommand\spro{{\em s}-process}
\newcommand\msol{\mbox{$M_\odot$}}
\newcommand\rbar{\mbox{$\bar{R}$}}
\newcommand\nsn{\mbox{$N_{\rm SN}$}}

\newcommand\Beta{{\rm B}}

\newcommand\avg[1]{\hbox{$\langle #1 \rangle$}}

\newcommand\beq{\begin{equation}}

\newcommand\eeq{\end{equation}}

\newcommand\fun[2]{\lower3.6pt\vbox{\baselineskip0pt\lineskip.9pt
  \ialign{$\mathsurround=0pt#1\hfil##\hfil$\crcr#2\crcr\sim\crcr}}}
\newcommand\pref[1]{(\ref{#1})}

\begin{document}

\title{A SIMPLE MODEL FOR {\em r}-PROCESS SCATTER AND HALO EVOLUTION}

\author{Brian D. Fields}
\affil{Center for Theoretical Astrophysics, Department of Astronomy, 
University of Illinois,
Urbana, IL 61801, USA}

\author{James W. Truran}
\affil{Department of Astronomy and Astrophysics and Enrico Fermi Institute, 
University of Chicago,
Chicago, IL 60637, USA}

\author{John J. Cowan}
\affil{Department of Physics and Astronomy, University of Oklahoma,
Norman, OK 73019, USA}

\begin{abstract}
Recent observations of heavy elements produced by
rapid neutron capture ({\em r}-process) in the halo have
shown a striking and unexpected behavior:
within a single star, the relative abundances of {\em r}-process
elements heavier than Eu are the same
as the same as those of solar system matter, while 
across stars with similar metallicity Fe/H, the
{\em r}/Fe ratio varies over two orders of magnitude.
In this paper we present a simple analytic model 
which describes a star's abundances in terms of 
its ``ancestry,'' i.e., the number of 
nucleosynthesis events (e.g., supernova explosions) which
contributed to the star's composition.
This model leads to a very simple analytic 
expression for the abundance scatter versus Fe/H,
which is in good agreement with the data and with
more sophisticated numerical models.
We investigate two classes of 
scenarios for {\em r}-process nucleosynthesis,
one in which {\em r}-process synthesis events occur in 
only $\sim 4\%$ of supernovae but iron synthesis
is ubiquitous, and one in which
iron nucleosynthesis occurs in only about 9\% of supernovae.
(the Wasserburg- Qian model).
We find that the predictions in these
scenarios are similar for 
$[{\rm Fe/H}] \ga -2.5$, but that these models can
be readily distinguished observationally by measuring the dispersion
in {\em r}/Fe at $[{\rm Fe/H}] \la -3$.
\end{abstract}


\section{Introduction}

Neutron capture processes dominate the nucleosynthesis of
elements beyond the iron peak.
The physics of two neutron capture mechanisms has
long been understood, and the astrophysical site for 
slow neutron capture nucleosynthesis (the \spro)
has been shown to be low-mass stars (e.g., Busso, Gallino, \& Wasserburg \cite{bgw}).
However, the site for
rapid neutron capture nucleosynthesis (the \rpro)
has not yet been unambiguously identified, although
it is very likely connected to massive stars.
The production site itself might be found
in some or all Type II supernova (SN) events
(e.g., Woosley, Wilson, Mathews, Hoffman, \& Meyer \cite{wwmhm}), or
in binary neutron star mergers
(e.g., Eichler, Livio, Piran, \& Schramm \cite{elps};
Rosswog, Davies, Thielemann, \& Piran \cite{rdtp}).
Truran \cite{jim81} showed that the halo stars
are a particularly useful laboratory for study of 
the \rpro, as the observed neutron capture abundance patterns 
in these stars indicate that the \spro\ component
drops out, and the \rpro\ dominates, as one 
looks at stars with $[\feh] \la -2.2$
(Burris et al.\ \cite{burris}).
Recent years have shown that halo stars indeed
give unique insight into the \rpro.
With the advent of high dispersion, high S/N
measurements, the abundance observations within and
among halo stars has led to surprising new discoveries
with important consequences for theories of the \rpro\ as
well as halo formation.

The halo star with perhaps the most striking \rpro\ composition
is CS 22982-052.  With $[{\rm Fe/H}] = -3.1$, this is an
ultra-metal-poor star;\footnote{
In the standard notation, 
$[A/B] = \log(A/B)_\star - \log(A/B)_\odot$;
we take ultra-metal-poor stars to be those with
$[{\rm Fe/H}] < -2.5$.
}
its abundance ratios through
the iron peak are typical for a halo star
However, this star also has been
observed in 20 \rpro\ elements from barium upwards (the ``heavy'' \rpro),
with abundances (Sneden et al.\ \cite{sneden96,sneden2000})
\begin{equation}
\label{eq:hir}
\left[ \frac{r}{\rm Fe} \right] \simeq 1.7 
  \ \ \Rightarrow \ \  
  \frac{r}{\rm Fe} \simeq 50 \left( \frac{r}{\rm Fe} \right)_\odot
\end{equation}
The element-to-element scatter in this ratio is consistent with
the observational errors.
This remarkable trend does {\em not} appear to hold for the
``light'' \rpro, $40 < Z < 56$ (Sneden et al.\ \cite{sneden2000}).
Nevertheless, the fact that the heavy \rpro\ elements agree so well
with each other is stunning, given that this 
star could well have abundances that reflect
the nucleosynthesis of a single supernova,
while the solar abundances average over 
many generations of supernovae.
The strong implication is that there is a {\em unique}
astrophysical site for the \rpro\ (at least for
the heavy $r$-elements, which for brevity we will refer to 
simply as ``$r$-elements'' hereafter).
 
This result has profound implications.
First, the mere presence of \rpro\ in very metal-poor
stars points to origins associated with massive stars.
Moreover, given the apparent {\em universality} of the
\rpro, one might expect that all ratios among \rpro\ nuclides to always
be constant (with pollution by the $s$-process increasing
with [Fe/H]).
This leads to the prediction that at the lowest metallicities, when
the $s$-process has not turned on, ratios among \rpro\ 
elements should be fixed and the same across all stars.
So far this appears to be true in the other ultra-metal-poor star
for which there is similarly good data, HD 115444
(Westin, Sneden, Gustafsson, \& Cowan \cite{wsgc}),
which has $[{\rm Fe/H}]=-3.0$ and $[\rfe] = 0.96$.
Very recently, Cayrel et al.\ \cite{cayrel}
have shown that CS 31082-001 (with $[\feh]=-2.91$) has Os/Fe and Ir/Fe
abundances which give $[\rfe] = 1.98$ and $R = 96$.
This would be the highest \rpro\ overabundance yet observed.
Clearly, a systematic study of \rpro\ abundance patterns in this
star is of the highest priority (particularly since this is the
first star in which uranium has been detected).

CS 22892-052 and HD 115444 have
iron abundances that are  identical within errors, 
and both show
a remarkable constancy in their $\rfe$ ratios for
different {\em r}-elements {\em within each star}.
However, these stars show a significant {\em difference} in 
their mean \rfe\ ratios.  This is an example of the large scatter in
\rfe\ ratios that has been observed in halo stars.
As we will see, the dispersion in \rfe\ ratios is small
at $[{\rm Fe/H}] \ga -1$, but grows with decreasing metallicity,
finally spanning two orders
of magnitude for ultra-metal-poor stars.
The observed large scatter in \rfe\ in
Pop II stars --- and the very high \rfe\ in a few --- demands that 
{\em not all stars make both the \rpro\ and iron.}

Models for \rpro\ nucleosynthesis and for Pop II chemical evolution
must account at once for the remarkable constancy of \rfe\
among different elements
within individual stars, as well as for the variability of \rfe\
between stars.
These facts together suggest the following simple picture for Pop II
\rfe.  
We assume that the observed Pop II \rfe\ abundances
in each star reflect \rpro\ contributions of
a {\em few} supernovae.
The basic idea is that 
\begin{itemize}
\item
Different \rfe\ arise from different mixing between the dominant 
\rpro\ sources
and the dominant Fe sources; and

\item
The \rfe\ scatter then reflects the amount of mixing between the two sources.
\end{itemize}
The key is the inhomogeneity of the halo, which
this scheme in fact quantifies.

This scenario makes quantitative predictions:
\begin{enumerate}
\item
There is a maximum to \rfe, which is the \rfe\ production ratio
in those stars which produce \rpro\ elements, 
undiluted by
any ({\em r}-poor) Fe events

\item
The minimum to \rfe\ is the \rfe\ ratio in 
the {\em r}-poor events.  This minimum may
be as small as $\rfe = 0$.

\item
The placement of a star between these extremes quantifies
a admixture of nucleosynthesis sources recorded in the star.  Thus, 
(a) at the very earliest times, when single SN events are really all
that contaminate a given star, the \rfe\ scatter should {\em only} populate 
the extremes.
(b) At later times, but before mixing is efficient, the regions
between the extremes will be filled in.
(c) At still later times, mixing becomes efficient, and then
the scatter decreases towards a "universal" mix of the two sources.
\end{enumerate}
If this line of reasoning is correct, it has the very
significant implication that 
\rfe\ becomes a {\em tracer} of the inhomogeneity of the halo.
One can see the transition from single to multiple events, and also
the transition from multiple sources to a well mixed Galaxy.
The details of these transitions allow one to correlate
metallicity and mixing.  And we can at least grossly confirm
the time sequence of the nucleosynthesis events. 

Several groups have used similar arguments to
motivate detailed models which explain
\rfe\ scatter in terms of {\em r}-process
nucleosynthesis and an inhomogeneous chemical evolution of the
Galactic halo.  Ishimaru \& Wanajo \cite{iw} 
construct a one-zone model for the halo, 
but introduce abundance dispersion through
a Monte Carlo realization of individual supernova
events with yields of Fe and Eu which are
strong function of mass.  They obtain good fits
to the observed scatter for models
in which the \rpro\ is produced by a small 
fraction of supernovae.
Argast, Samland, Gerhard, \& Thielemann \cite{asgt}
use similar \rpro\ progenitor masses, but
drop the assumption of one-zone evolution.
They model the spatial inhomogeneity and incomplete mixing
of expanding supernova remnants in the halo,
and again find good agreement with the observed Eu/Fe
dispersion.

The goal of the present paper is to aid in the understanding
of these detailed models, and to help focus attention on
the key physics relevant for this problem.
We will show (\S \ref{sect:models}) that the \rfe\ scatter follows quite
generally from the stochastic behavior of the halo
chemical evolution, and that one can derive simple, explicit, analytical
expressions for the abundance scatter as a function of
metallicity.  We will do this for two scenarios,
one in which the \rpro\ occurs in only a small 
fraction of supernovae but Fe is ubiquitous (\S \ref{sect:Fe-dom}), and 
the scenario of Wasserburg \& Qian \cite{wq}, 
in which the converse holds (\S \ref{sect:WQ}).
Fortunately, these scenarios can be distinguished 
by future measurements of \rfe\ scatter in ultra-metal-poor stars,
and we present observational strategies to do this
(\S \ref{sect:tests}).
Discussion and conclusions appear in \S \ref{sect:conclude}.

\section{Two Scenarios for \rpro\ Origin}
\label{sect:models}

The \rfe\ scatter demands that
the \rpro\ and iron are not co-produced in the
same ratios in the same events, but
rather there will be {\em r}-rich and {\em r}-poor events.
Also, very high and very low \rfe\ ratios are observed.
The highest \rfe\ event observed places a lower limit 
on the \rfe\ yields of the {\em r}-rich events.
Furthermore, the fact that the highest \rfe\ is
much larger than the Pop II mean implies that
the {\em r}-rich, high \rfe\ events occur
at a very different rate than low \rfe\ events.
One can envision either ``iron-dominated'' schemes
in which high \rfe\ events are rare, but iron-producing events
are not; or, alternatively, the possibility that
\rpro\ events are the more common, with iron-producing events
rarer.  We now consider each of these scenarios in turn.

\subsection{Iron-dominated Models}
\label{sect:Fe-dom}

We assume (1) that the supernova \rpro\ yields are
{\em bimodal}, i.e., that some supernovae have a high
\rpro\ yield and others a low one (possibly zero);
and (2) for simplicity, we assume not only that all supernova make
iron, but also that the Fe yield takes the same constant
value for all supernovae.\footnote{It is possible that some
supernova make neither $r$ nor Fe; if such events exist,
they are irrelevant to the present discussion.}
With these assumptions, supernova ejecta can take one
of two values of \rfe.
We will denote the objects with the larger
\rfe\ ratio as class $A$, and those with the lower
ratio as class $B$.  We purposely avoid
a more descriptive nomenclature--e.g., ``high'' and ``low''--as
we wish to avoid both confusion (high yields? rates? masses?) 
and prejudice with regard
to the nature of the progenitors which are the dominant
$r$ producers.
To simplify notation, we will define the scaled \rfe\
ratio to be 
\beq
R \equiv \frac{\rfe}{(\rfe)_\odot}
\eeq
which implies that $[\rfe] = \log R$.

Thus, a supernova will produce a fixed amount of iron,
and yields {\em r}-elements at either the high value $R_a$ or the 
low value $R_b$.
Denote the fraction of class $A$ supernovae (averaged over the halo IMF)
to be $f_a$; then the fraction of class $B$ supernovae
is of course $f_b = 1 - f_a$.

The \rfe\ data show that at a metallicity
$\feh \sim -1$, $[\rfe]$ converges to
a value of $\sim 0.3$, i.e., $\rbar \sim 2$.\footnote{Above 
this metallicity \rfe\ decreases, but this
is presumably due to the addition of the Fe
yields of Type Ia supernovae.  We do not include these
events (and implicitly assume their \rpro\ yield
to be zero); thus we will not attempt to model
\rfe\ at $[\feh] > -1$.}  
We take this value as the mean $R$ attained when
averaging over a large sample of progenitors,
whose fractions reflect the underlying
$f_a$ and $f_b$ dictated by the IMF.
Thus we have that 
\beq
\label{eq:rbar}
\rbar = f_a \; R_a + f_b \; R_b = R_b + f_a \; \left( R_a - R_b \right)
\eeq
We can use the observed scatter to estimate 
the values of $R_a$ and $R_b$.  The highest
observed is $[\rfe]\simeq 1.7$ (eq.\ \ref{eq:hir}); we take this as a 
measure of $R_a \simeq 50$.  On the other hand,
the lower limit of the \rfe\ scatter is not
as well determined.  
The available data 
suggests that the minimum, if any,
differs depending on the elemental {\em r}-tracer used.
At most, we have $[\rfe]_{\rm min} \sim -0.6$, giving
$R_b \simeq 0.25$; it is also possible that
$[\rfe]_{\rm min} \sim -1.7$, giving
$R_b = 0.02 \simeq 0$.  
Upper limits
to \rfe\ have not been reported, but this
may be a selection effect.  A key issue is whether there
{\em could have been} lower \rfe\ values than
the ones reported; such information places useful
constraints on \rpro\ production.  
For our purposes, as long as $R_b \ll \rbar$, then
we may take $R_b = 0$, as we will now see.

Once one establishes the observed values for $\rbar$, $R_a$, and $R_b$,
then the relative numbers of class $A$ and $B$ supernovae is
now {\em fixed}.  Eq.\ \pref{eq:rbar} gives 
\beq
\label{eq:fa}
f_a = \frac{\rbar - R_b}{R_a - R_b} \simeq \frac{\rbar}{R_a}
\eeq
where the last expression uses the observed fact that
$R_b/\rbar \ll 1$.
For $\rbar = 2$ and $R_a = 50$,
eq.\ \pref{eq:fa} gives 
$f_a = 3.5\% $ for  $R_b = 0.25$,
and $f_a = 4.0\% $ for  $R_b = 0$.
Thus we see that observed fact of the very large \rfe\ variations
immediately implies that class $A$ events are
{\em required} to be uncommon, regardless of 
their physical nature.
That is, {\em the events which produce the \rpro\
originate from a small fraction, $\sim 4$\%, of all massive stars}.

\subsubsection{Basic Model}

Using these numbers, we are now in the position
to model the scatter of \rfe.
To do this, we create a set of halo stars
and deduce the history--the nucleosynthetic ancestry--of each.
We assume that each of the halo stars we create
incorporates gas which has been enriched by
some number $\nsn$ of supernova of either
type:  this is the number of supernova
ancestors for the star.
We create stars by allow $\nsn$ to run from 1
to $N_{\rm max}$, and sample equal intervals
in $\nsn$.  Note that we are free to chose the number
of halo stars we create, and that this number is unrelated to the number of
supernova ancestors assigned to a given halo star.

The metallicity follows from the assumption that
all stars have the same iron yield, so that
$\feh \propto \nsn$.
We introduce a parameter $(\feh)_{\rm min}$, the
iron enrichment from a single star.
Then $\feh = \nsn (\feh)_{\rm min}$,
and $[\feh] = [\feh]_{\rm min} + \log \nsn$.

The \rpro\ pattern of the stars is also fixed 
in terms of the stellar ancestry.
We assume that the mean value of $R$ 
for our population of stars is always $\rbar$.\footnote{As long
as both classes $A$ and $B$ exist; below we will consider the
effect when there is a time delay before the onset of one class.
The constancy of \rfe\ can be determined observationally.
Current data samples the low metallicities too sparsely
to make a strong statement (Burris et al.\ \cite{burris}), 
but with more observations
one should be able to test for a metallicity dependence
of \rfe.}
We then introduce scatter around $\rbar$ by
sampling stars whose ancestry (in class
$A$ and $B$ stars) is a random variable.
The underlying physics of this random process is
ultimately controlled by the physics of
\rpro\ nucleosynthesis and the mixing of
stellar ejecta.  However, even in the absence of
this detailed physics, simple assumptions
allow for a stochastic behavior that resembles
the observations.  We will illustrate two such
simple assumptions regarding the random distribution of 
``ancestors.''

First, we treat $N_a$, the numbers of ancestors of class $A$,
as a {\em discrete} random variable.
That is, we imagine that the number 
ancestors from each class take integer values.
We choose $N_a$ from a binomial distribution
with mean $\bar{N}_a = f_a \nsn$;
it then follows that $N_b = \nsn - N_a$.
For each star, the fraction of class $A$ (high {\em r})
progenitors is 
$f_{a,\star} = N_a/\nsn$, and
\beq
\label{eq:rstar}
R_{\star} = f_{a,\star} R_a + f_{b,\star} R_b 
   = R_b + (R_a - R_b) \, N_a/\nsn
\eeq
The dispersion in $N_a$ naturally induces a
dispersion in $R_\star$:
$\sigma(R_\star) = (R_a - R_b) \, \sigma(N_a)/\nsn$
Since $\sigma(N_a) \sim \sqrt{N_a}$, this gives
\beq
\label{eq:sigR}
\sigma(R_\star) \simeq (R_a - R_b) \, \sqrt{N_a}/\nsn
  \simeq \rbar/\sqrt{N_a}
  = (f_a \, {\rm Fe/Fe_{min}})^{-1/2} \ \rbar
\eeq
where we have assumed successively that
$f_b \simeq 1$ and $R_b \ll \rbar$.
Thus we see that the fractional error in $R$
is just given by the counting statistics of $N_a$:
$\sigma(R_\star)/\bar{R}_\star \simeq 1/\sqrt{N_a}$.
We now can put each star on the (\rfe,$[\feh]$) plane, and
the model is complete.

This model thus has three free parameters which can
be fixed by observations:   $\rbar$, $R_a$, and $R_b$;
there is one parameter, ${\rm (Fe/H)_{min}}$ which is (for now) 
determined by theory.
We can rewrite eq.\ \pref{eq:sigR}
explicitly in terms of these parameters:
\beq
\label{eq:env}
\sigma_R = (f_a \, {\rm Fe/Fe_{min}})^{-1/2} \ \rbar
\eeq
and thus
\beq
\label{eq:logenv}
\sigma_{[r/{\rm Fe}]} = \frac{1}{f_a^{1/2} \ln 10} \ 
    10^{-({\rm [Fe/H] - [Fe/H]_{min}})/2}
 = 10^{-({\rm [Fe/H]} + 3.3)/2}
\eeq
The model thus predicts that the intrinsic
scatter in \rfe\ falls as $(\feh)^{-1/2}$.
As we will see, this is indeed consistent with the halo star data.
Of course, the data also contain observational errors, both in
[\rfe] and [\feh].  We include these in the model, using
$\sigma([\rfe])_{\rm obs} = 0.10$ dex, and 
$\sigma([\feh])_{\rm obs} = 0.05$ dex, with the 
supposition that
the errors are uncorrelated with the intrinsic dispersion.

\begin{figure}
\epsscale{0.7}
\plotone{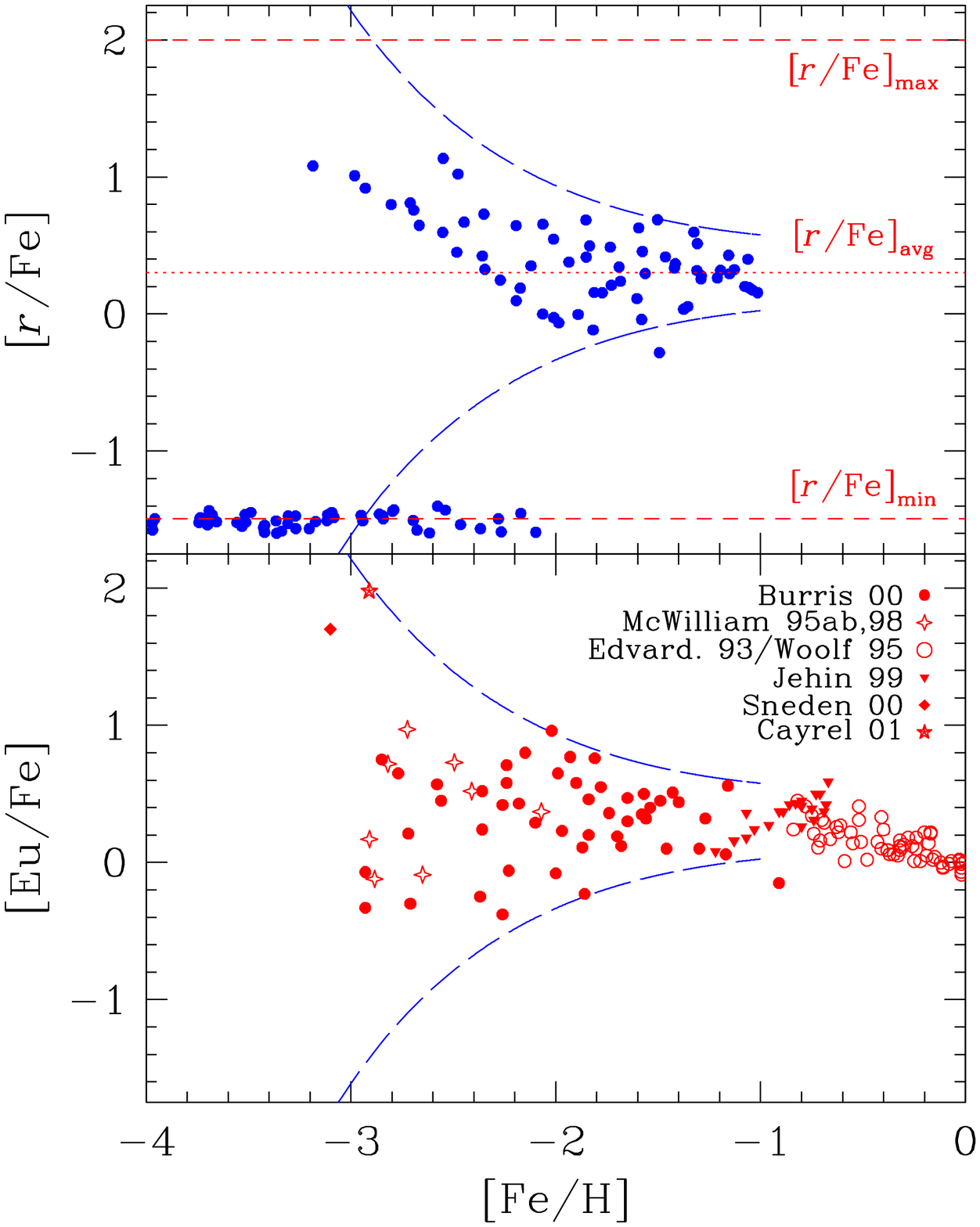}
\caption{
Distribution of \rfe\ abundances for halo stars.
{\em Upper panel}: model calculation.  The points represent
stars whose ancestry has been drawn according to a binomial
distribution.  In addition to the intrinsic dispersion
calculated in the model, we have included random observational
errors with zero mean and standard deviation
$\sigma_{[R]} = 0.15$ and $\sigma_{\rm [Fe/H]} = 0.10$.
{\em Lower panel}: observed data for
[Eu/Fe]; references are as shown.  
For CS 31082-001 (Cayrel et al.\ \cite{cayrel})
Eu abundances are as yet unavailable, so the average of
[Os/Fe] and [Ir/Fe] is used.
The dashed curves in both panels are the analytic relation for the
$2-\sigma$ envelope predicted from this model (eq.\ \ref{eq:env}). 
The curves are drawn to account for the logarithmic nature of the
plot, ie., the curves are located at the $2\sigma$ limits, namely
$\log \rbar \pm 2 \log 10 \sigma(R)/\rbar$.
\label{fig:binomial}
}
\end{figure}

Results for a Monte Carlo realization of
a population drawn from a binomial distribution appear in Figure 
\ref{fig:binomial}.
The predicted envelope  is shown, and
provides a reasonable fit for the observed data.
Note that the envelope includes both the observational
errors as well as the intrinsic scatter 
given by (eq.\ \ref{eq:env}).
In the binomial model just described,
there are quantization effects due to the
integral values of $N_i$.  These are
particularly noticeable when
$\bar{N}_a < 1$, which occurs when
$\nsn < 1/f_a \simeq 25$, or $[\feh] \la -2.6$.
In this regime, most points will cluster at 
$R_b$, with only a few at higher values.
Furthermore, $R = \rbar N_a/\nsn$ is also
quantized. 
The lower
envelope of the nonzero values is given by 
the case when $N_a = 1$, so that $R_{\rm env} = \rbar/\nsn$
and  $[\rfe]_{\rm env} = -\feh + \log \rbar$;
this is clearly seen in Figure \ref{fig:binomial}.

One need not require that a star's heavy element ancestry
be quantized in this way.
For example, one can plausibly imagine that the mixing of
successive generations of supernova ejecta is not
an ``all or nothing'' prospect, but rather that
a given parcel of ISM gas and dust can be enriched to different
degrees by the ancestors it had.
This is indeed very likely to be the case, 
which would mean that the quantization effects of the binomial model
are spurious (and thus the binomial results are only to be trusted
in the regime where $\bar{N}_a > 1$, or $[\feh] \ga -2.6$.

Thus, it is of interest to consider making the number of ancestors
of each class a continuous random variable.
This can be done in a way that naturally generalizes
from the binomial distribution; in that case,
 the total number of supernova progenitors $N_{\rm SN}$ is fixed, and 
the {\em fraction} $f = N_a/N_{\rm SN}$ of progenitors of class $A$
is a random variable which has a mean $\avg{f} = f_a$
but can take only discrete rational
values due to the integral nature of $N-a$.  
We will continue to take
$\nsn$ to be fixed, but we
will now assume that the fraction $f$ of progenitors
of class $A$ is now a continuous random variable, with mean
$\avg{f} = f_a$.  We thus want to draw $f$ from a 
continuous distribution
with values in the interval $[0,1]$ and with a fixed mean.
These requirements are met by the beta distribution,
which has a distribution function
\beq
\xi(x) = \frac{1}{\Beta(a,b)} \  x^{a-1} \ (1-x)^{b-1}
\eeq
for $f \in [0,1]$ and 
where $\Beta(a,b)$ is the beta function, which can be
expressed in terms of the gamma function 
$\Beta(a,b) = \Gamma(a)\Gamma(b)/\Gamma(a+b)$.
We fix the parameters $a$ and $b$ of the distribution 
by simultaneously fitting the mean $\avg{f} = a/(a+b)$
to the observed value $f_a$, and forcing the
variance $\sigma^2(f) = ab(a+b)^{-2}(a+b+1)^{-1}$
to the same value as the binomial
case, $\sigma^2(f) = f_a (1 - f_a) / \nsn$.
These conditions are satisfied if
$a = f_a (\nsn -1)$ and $b = (1-f_a)(\nsn-1)$.

\begin{figure}
\plotone{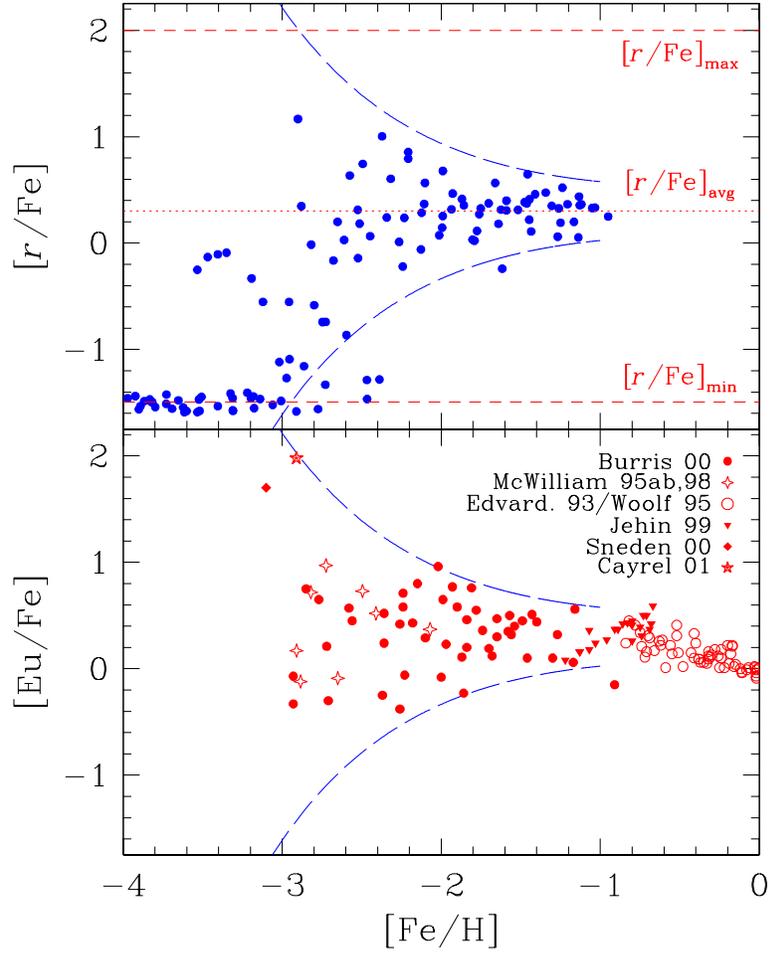}
\caption{
As in Figure \ref{fig:binomial}, for
the beta distribution model.
Due to the continuous nature of the parent distribution,
the quantization effects seen in the binomial case
are now absent.  
\label{fig:betadis}
}
\end{figure}

Our procedure is to draw $f$ according to the beta distribution
$\xi(f)$.  We then find $N_a = f \nsn$ and $N_b = (1-f) \nsn = \nsn - N_a$.
Since $f$ is a continuous variable, so are $N_a$ and $N_b$.
Our choices
of the parameters $a$ and $b$ guarantee that 
$\avg{R} = \rbar$ and 
that $\sigma(R)$ is identical to the value found for the binomial case,
so that eqs.\ \pref{eq:sigR} and \pref{eq:env} still hold.
Results for a Monte Carlo simulation of a
stellar population appear in Figure \ref{fig:betadis}.
We see that the general trend is quite similar to 
that for the binomial case, but that the continuous
nature of the parent distribution guarantees that 
quantization effects are now absent.
We see
that at the very lowest metallicities $[\feh] \la -3.5$, 
the points cluster at $\rfe_{\rm min}$, 
and have a mean value below that at higher metallicity.
This feature has as similar origin to that of its counterpart
in Figure \ref{fig:binomial}.
At the lowest metallicities, 
the small number statistics 
in the SN parents is a dominant effect, and
the {\em r}-rich events are too rare to be 
expected in a sample of the size presented here.

A comparison with the observed data in Figure \ref{fig:betadis}
shows that the beta distribution model gives a good
fit to the available data.
We thus conclude that the observed scatter in \rfe\
may be understood by our simple scenario in which
halo stars stochastically sample the yields of
two different populations of heavy element producing events.
In this interpretation, the degree of scatter
increases with decreasing metallicity due to
counting statistics, so that 
the lowest-metallicity events record the nucleosynthesis of
a few ($\nsn = {\rm Fe/Fe_{min}}$) events.
Furthermore, since 
$\sigma(R)/\rbar \simeq \sqrt{f_a}/\nsn$, 
a good measure of the scatter in \rfe\ 
allows one to find $\nsn$ and thus one can fix
the parameter ${\rm (Fe/H)_{min}}$. 
In this way, one can empirically 
measure the single-event iron yield,
and also determine the number of nucleosynthesis
events recorded at a particular metallicity.
The fact that the observed distribution is
reasonably enclosed by the theory curves 
which use 
$[\feh]_{\rm min} = -4$ 
(e.g., Audouze \& Silk \cite{as})
indicates that this
value is a reasonable first approximation.

\subsubsection{Introducing Stellar Lifetimes}

Having achieved a reasonable fit to the data, we
are emboldened to consider refinements to this model.
Specifically, we turn to the question of timescales.
Thus far, we have implicitly assumed that
there is always the same relative probability
of {\em r}-rich and {\em r}-poor 
nucleosynthesis events.
However, this assumption is likely to break down
at very early times.  One expects that 
the to classes $A$ and $B$ of nucleosynthesis events
stem from different physics.  For example, the
{\em r}-rich and -poor classes could correspond to
progenitor stars with different masses and thus
different lifetimes, or
\rpro\ production could arise from binary neutron star mergers,
and thus require some time delay for inspiral.
In either case, one expects a time lag between the first
events of one class versus the first events of the other.
During this initial period, the \rfe\ ratio of the ISM
and of any new stars will be only that of the allowed
class of nucleosynthesis events.  There will thus be
no dispersion in \rfe\, or rather, the dispersion will be
due to the smaller intrinsic dispersion within the allowed
class.  
Once events of the other class
can occur, \rfe\ dispersion will set in, and the
\rfe\ will scatter around its mean value.

We can crudely simulate this time delay between the
onset of the two classes as follows.  
We note that the iron abundance increases with time.
Thus, a time delay can be encoded as a delay in
\feh.  We choose a value $(\feh)_{\rm cut}$ 
as the cutoff that marks the 
onset of the lagged class of nucleosynthesis events.
To allow for dispersion in the birth times
of the first stars, we introduce some randomness
in this cutoff by making the cutoff number ${N}_{\rm cut}$ of
events a poisson random variable, with mean
$\bar{N}_{\rm cut} = {\rm Fe}_{\rm cut}/{\rm Fe}_{\rm min}$.
We then require that if $\nsn < N_{\rm cut}$, only
events of the allowed class $i$ can occur,
and thus $R = R_i$.  If $\nsn > N_{\rm cut}$, then
we follow the standard procedure described above.

\begin{figure}
\plotone{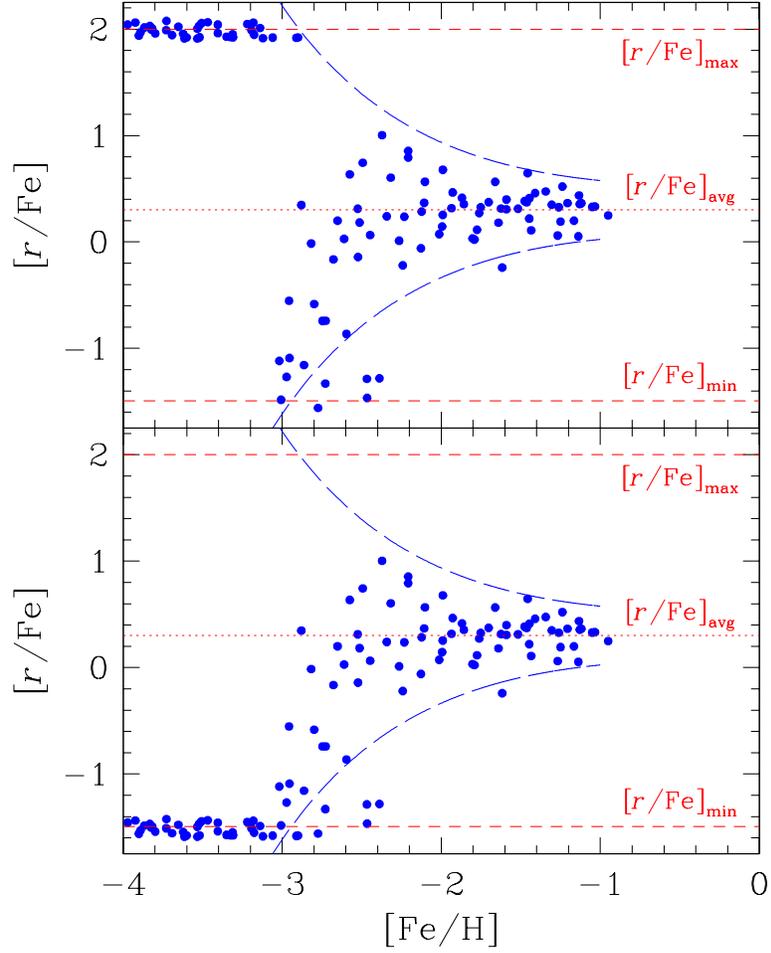}
\caption{
Predicted \rfe\ distributions, with timescale information
encoded through metallicity cutoffs.
{\em Upper panel}:  a delay has been imposed on class $B$
({\em r}-poor) events, so that class $A$ events dominate at early
times.  {\em Lower panel}:  a delay has been imposed on class $A$
events.
\label{fig:cutoffs}
}
\end{figure}

Results appear in Figure \ref{fig:cutoffs}, using
${\rm Fe}_{\rm cut} = 10 \; {\rm Fe}_{\rm min}$,
which roughly corresponds to timescales
$\tau_{\rm long} = 10 \; \tau_{\rm short}$,
where the longer lifetime is that of the species cut off.
As expected, we see that the very low $\feh$,
the dispersion vanishes and the points lie
at the level of the allowed class.
While this is clearly an oversimplified description,
it nevertheless gives a qualitative sense of the kind of 
behavior one expects if there is indeed a significant disparity
between the timescales for the production of 
{\em r}-rich and -poor events.  
A discrepancy in timescales is required if the \rpro\ 
production ``switch'' is related the progenitor star's mass.
As we suspect the mass to be the controlling parameter,
our prediction is that one of the panels
in Figure \ref{fig:cutoffs} represents the trend
that will be observed in ultra-metal-poor stars.

Thus, a discovery
of the sort of behavior seen in Figure \ref{fig:cutoffs}
would indicate which class has the sorter production timescale,
and thus provide a key clue as to the astrophysical origin of
the dominant {\em r}-process site.  Furthermore, a determination
of ${\rm Fe}_{\rm cut}/{\rm Fe}_{\rm min}$ would help to
quantify (in a model-dependent way)  the time delay itself.
If class $A$ stars have the longer timescale,
$\tau_a > \tau_b$,
then physically this means that
(1) the \rpro\ producers evolve more slowly and
thus (2) these stars should show up later, so
that there should be a limiting low metallicity
below which all stars have $R_\star = R_b$.
The appearance of the first stars with 
$R_\star > R_b$ marks the time of the first \rpro\ stars
and thus $t \ge \tau_a$.  

We can be more specific by using the fact that \rpro\ stars
must be a small fraction of all supernovae.  This means
that the progenitors must represent a limited range in the 
massive star IMF:  either $8-10 \msol$ or $\ga 25 \msol$.
If $\tau_a$ is long, this would select the $8-10 \msol$
range, and thus mean that 
$\tau_a \simeq \tau(10\msol) \simeq 3 \times 10^7$ yr.
Now consider the opposite case.  If $A$ stars have the shorter
timescale, the $\tau_a < \tau_b$, and 
the \rpro\ stars should appear first.
Consequently, there should be a limiting low metallicity 
below which all stars have
$R_\star = R_a$, and the appearance of the first stars
with $R_\star < R_a$ marks the time of the first 
non-\rpro\ stars and thus $t \ge \tau_b$. 
The short $\tau_a$ implies the high mass
($m \ga 25 \msol$ )progenitor
mass range, and thus $\tau_b \ga \tau(25\msol) \simeq 10^7$ yr.

While massive stars seem required to explain the 
\rpro\ simply due to their evolutionary timescale,
it is possible that the \rpro\ is not made in the 
core collapse and explosion, but rather through
the merger of supernova remnants, via
neutron star-neutron star mergers 
(e.g., Eichler, Livio, Piran, \& Schramm \cite{elps};
Rosswog, Davies, Thielemann, \& Piran \cite{rdtp}).
As pointed out recently by Qian \cite{qian},
an important constraint on this scenario
comes from magnitude of the halo star \rfe\ scatter,
which appears to be too small (!) to reconcile with
the low merger rate and thus high yields of the 
coalescence events.
An additional
issue for neutron star coalescence is the 
timescale of the merger, which is an extra delay
in the appearance of the \rpro\ compared to the
other supernova products (such as iron).
This timescale is a strong function of
the orbital semimajor axis and particularly eccentricity
(Peters \cite{peters}).  
As shown in population synthesis calculations (e.g., Fryer, Woosley,
\& Hartmann \cite{fwh}), this leads to a wide distribution
of inspiral timescales, spanning a few Myr to $\sim 1$ Gyr.
Thus, if the \rpro\ does come from binary neutron star 
mergers, then the first events had to have the correct
orbital parameters to allow a very rapid evolution.

Finally, even if observations establish the {\em lack} of a signature of the
kind seen in Figure \ref{fig:cutoffs} would provide useful
information about the \rpro\ as well.
We would learn that there is no significant difference in the
timescales for production of the two classes of events.
This would rule out any models which require such a delay,
and would challenge the presumption that the \rpro\ yields
are a function of the mass of the progenitor stars.
Alternatively, it may suggest a very different
scenario entirely, as we now shall see.

\subsection{{\it r}-process-dominated Models}
\label{sect:WQ}

In the previous section, we assumed that \rpro\ nucleosynthesis
events are rare, while iron production events are ubiquitous
(for massive stars).  We now turn to the reverse case, 
which has been developed in some 
detail by Wasserburg \& Qian (\cite{wq}; hereafter, WQ2000).
The essential ingredients are the following,
in chronological order:
(1) ``Pop III.'' A very rapid burst of star formation
produces an uneven Galactic floor of $[\feh]\sim -4$ to $-3$,
but no \rpro.  The inhomogeneity of this population will
lead to \rfe\ scatter in this metallicity range.
(2) ``Pure \rpro.''  The first of the Pop II supernovae
produce the (heavy) \rpro\ without any Fe.
These are WQ2000's high-frequency {\it H} events.
(3) ``Fe added.''  After $\sim 10$ {\em H} lifetimes, 
the first iron-producing supernovae go off.  These make
the light \rpro\ (i.e., sub-barium, which we do not consider here),
but {\em none} of the heavy \rpro.
They occur at lower frequency and thus are dubbed {\it L} events.

We model the 
Wasserburg \& Qian scenario as follows; our approach
and some details are very similar to the
model discussed in the previous section.
WQ2000 note that the Pop III events need not occur
in a homogeneous or well-mixed way, which will
lead to global [\feh] variations between $-4$ and 
about $-3$.
Thus, we assume that a single Pop III parent
would have ejecta which mix with it surroundings,
leading to an enrichments of $[\feh] = [\feh]_{\rm min,III} = -4$;
this is the Pop III iron yield.
This sets the Fe floor.  These events do not make any \rpro,
which raises the possibility of an observable 
abundance record
of this stage remains, as we will discuss shortly.  
It is possible, indeed it is likely, that the first
stars of the next stage of Galactic evolution form prior to 
the onset of first {\it H} events which make the first
\rpro\ elements.  Any such stars will have only Pop III abundances
and thus will be void of any \rpro.  The number of such
{\em r}-free stars depends on the details of star formation
immediately after Pop III, and thus are difficult to estimate.
We will thus make the simple assumption that the
first {\em H} events are not significantly delayed after
the formation of the first halo stars, so that
there is not a significant population of Pop III, {\em r}-free
stars.

In the next stage,
the Pop II stars commence. 
The pure-$r$ {\it H} events occur first, as
they have lifetimes 10 times shorter
than those of the Fe-only {\it L} events.
Each {\it H} event is taken to produce the same \rpro\ yield,
and thus the total $r$-production is just the number
$N_h$ of parents times the yield of a single event.
Once the {\it L} events commence, they produce only Fe (and the
light \rpro, which we don't consider).  
WQ2000 put the Fe yield of a single event to
be about $[\feh]_\ell = -2.5$, i.e., 30 times more than that of
the Pop III events.  We assume all {\it L} events have
this yield.  

We thus have two epochs to model.  The first
is the period after the Pop III events but before
the first {\em L} event, and the second is the
rest of the Pop II phase, up to $[\feh] = -1$.
We begin modeling each halo star's ancestry
by determining which of these
epochs it samples.  To do this, we choose
a metallicity in the range $-4 \le [\feh] \le -1$.\footnote{Again, 
we are free to sample any metallicity in this
range, as we are not modeling the massive stars but only
the present-day halo stars which record the nucleosynthetic past.}
We note that 
whenever $[\feh] < [\feh]_\ell = -2.5$, then the star
must have arisen in the first, {\em H}-only epoch,
and have no {\em L} event progenitors.
In this case, the number of Pop III progenitors
is then fixed to be $N_{\rm III} = {\rm Fe/Fe_{min,III}}$.\footnote{This
may not be an integer, but we will interpret fractional progenitor
numbers as indicative of partial mixing, as done in the previous
section.}

We then determine the number of {\em H} events in our
halo star's lineage, by choosing $N_h$.  As in the previous
section, we note that the specific distribution of $N_h$ 
is determined by the detailed physics of star formation
and gas dynamics
in the earliest epoch in our Galaxy.  
The distribution is constrained by 
the {\em H} event-to-{\em L} event
ratio of 10, which demands that on average, there are 
10 {\em H} events before the first {\em L} event occurs.
Independent of the specific distribution, 
a crucial issue is the expected correlation
between $N_h$ and $N_{\rm III}$ (and thus Fe).
Since these events occur in two different phases
by two different population, we will make the simple assumption
that the two are uncorrelated.
We then have $R_\star = r_h N_h/{\rm Fe}$, with Fe 
already fixed
and with $r_h$ the \rpro\ yield of a single {\em H} event
(determined below).
Note that since $N_h$ is uncorrelated with
Fe, then the mean trend $\avg{R} = \avg{N_h} r_h/{\rm Fe}$
which scales as ${\rm Fe}^{-1}$.  That is, the
mean \rfe\ trend for $-4 \le [\feh] \le -2.5$ should
not be constant, but should decrease with logarithmic slope -1.
Scatter about the \rfe\ trend is created by
the variance in $N_h$:  $\sigma_R = \sigma(N_h) r_h/{\rm Fe} 
= \sigma(N_h)/N_h \; R$.
For illustration, we will choose 
$N_h$ from a uniform distribution.
in $(0,10)$, which gives $\avg{R} = 5 r_h/{\rm Fe}$,
and $\sigma_R/R = 1/\sqrt{3}$.

We now turn to the case of stars for which
we have picked a higher $\feh > \feh_\ell$.
For these stars, some contribution by
iron-producing {\em L} events is necessary.
As before we generate a number $N_{\rm SN}$ of supernova parents
for each halo star
Thus, each star generated has $N_h$  {\it H} parents
and $N_\ell$ {\it L} parents, 
but the star's iron abundance constrains
our choice of $N_\ell$ and thus 
$N_{\rm SN}$. 
Namely, the star has
a metallicity
\beq
\label{eq:FeWQ}
(\feh)_\star = N_{\rm III} (\feh)_{\rm min,III} + N_\ell (\feh)_\ell
  = \left( N_{\rm III,eff} + N_\ell \right) (\feh)_\ell
\eeq
with $N_{\rm III,eff} = ({\rm Fe}_{\rm min,III}/{\rm Fe}_\ell) N_{\rm III}$.
We choose $N_{\rm III}$ from a distribution
between 0 and ${\rm Fe}_\ell/{\rm Fe}_{\rm min,III} \simeq 30$;
for illustration, we will use the uniform distribution
Once given $N_{\rm III}$, then $N_\ell$ is fixed by
eq.\ \pref{eq:FeWQ}.

Turning to $N_h$, we must choose this in a way that is consistent
with our choice of $N_\ell$.  Note that since
the {\em L} events are 10 times rarer, the variance
in $N_\ell$ is much larger than that in $N_h$, and
thus we should expect a large span in the allowed
$N_h$ choice at fixed $N_\ell$.  We proceed by first estimating the
total number of supernova events to be $\hat{N}_{\rm SN} = N_\ell/f_\ell$,
where $f_\ell = 1/11 = 1 - f_h$ is the fraction of
supernova of type {\em L}.
Using a beta distribution, we choose $f_{\ell,\star}$,
with $a = f_\ell(\hat{N}_{\rm SN} -1)$ and $b = f_\ell(\hat{N}_{\rm SN} -1)$.
We then have $f_{h,\star} = 1 - f_{\ell,\star}$, and 
can determine $N_{\rm SN,\star} = N_\ell/f_{\ell,\star}$.
From these, we find
$N_h = f_{h,\star} N_{\rm SN,\star} = f_{h,\star}/f_\ell \ N_\ell$, and 
the star's \rfe\ ratio,
\beq
\label{eq:R_WQ}
R_\star = \frac{ N_h }{ N_{\rm III,eff} + N_\ell } \ R_{\rm yld}
\eeq
where $R_{\rm yld} = r_h/{\rm Fe}_\ell$ is the ratio of 
the (number) yields from the two event types.
Note that at high metallicity, we have $N_h, N_\ell \gg N_{\rm III,eff}$
and thus
\beq
\label{eq:WQlimit}
R_\star \longrightarrow \frac{ N_h }{ N_\ell } \ R_{\rm yld} 
\longrightarrow \rbar 
\eeq
But since Wasserburg \& Qian specify that H events
occur 10 times more frequently than L events,
we have $N_h/N_\ell = 10$, and from eq.\ (\ref{eq:WQlimit})
we can infer the proper $R_{\rm yld} = r_h/{\rm Fe}_\ell$ parameter
to give the correct $\rbar$.  Namely, we have
$R_{\rm yld} = \rbar/10 = 0.2$, and 
$r_h ={\rm  Fe}_\ell R_{\rm yld} = 6.3 \times 10^{-4}$.

We now turn to the \rfe\ dispersion at $\feh > \feh_\ell$.
Note that simple analytic
formulae are not easily derived for the \rfe\ scatter in the
WQ2000 scenario, owing to the presence of random variables
in both the numerator and denominator of eq.\ \pref{eq:R_WQ}.  
Nevertheless, one can roughly estimate that 
$\sigma(R)/R \sim 1/\sqrt{N_\ell}$.
For ${\rm Fe > Fe_\ell}$, this gives an intrinsic dispersion of
$\sigma(R)/R \sim \sqrt{\rm Fe/Fe_\ell}$,
the same ${\rm Fe}^{1/2}$ scaling as that in the Fe-dominated case.

\begin{figure}
\plotone{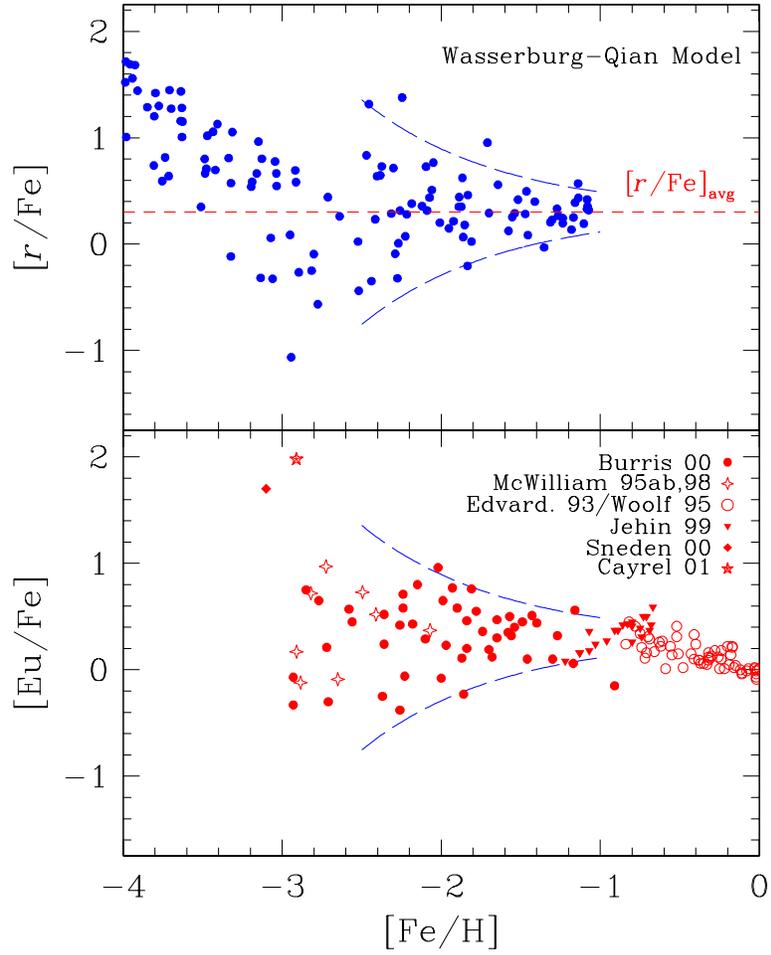}
\caption{
As in Figure \ref{fig:binomial}, for
the Wasserburg \& Qian \cite{wq} model.
\label{fig:WQ}
}
\end{figure}

Results from a Monte Carlo realization appear in Figure \ref{fig:WQ}.
the dashed curve shows the $2\sigma$ envelope which includes observational
errors and the intrinsic scatter predicted by
$\sigma(R)/R = \sqrt{\rm Fe/Fe_\ell}$ and plotted over its
range of validity.
We see that the scatter at $[\feh] > -2.5$ is similar to that in the 
Fe-dominated case, and in reasonable agreement with the 
present data.
However, the low metallicity convergence
behavior seen in Figure \ref{fig:cutoffs} does not appear in
the Wasserburg \& Qian scenario.  
Rather, at $[\feh] < -2.5$ we see the $R \sim 1/{\rm Fe}$ trend
discussed above.
We again conclude that
observations of the
\rfe\ scatter in this regime can thus provide important information
about \rpro\ astrophysics.

\section{Observational Tests of $r$-Process Astrophysics}
\label{sect:tests}

The models we have presented for \rfe\ scatter all
show similar trends from $\feh = -1$ down to
about $-2.5$.  However, below this metallicity,
the differences become increasingly stark, as seen
in the figures.  
The predicted behaviors are for points to cluster
above, below, or around the mean
value $[\rfe]_{\rm avg}$.
\begin{itemize}
\item
If the points all cluster around $[\rfe] \gg [\rfe]_{\rm avg}$,
this points to a very early population which, regardless
of the small number of ancestors, maintained high \rfe.
This would confirm that the earliest stars produced  both iron
and $r$-elements, and thus that the sources of the \rpro\
are short-lived and thus high-mass stars.

\item
If, conversely, the points all cluster around $[\rfe] \ll [\rfe]_{\rm avg}$
(or which have only upper limits for \rpro\ elements),
this indicates that the earliest stars produced iron but failed
to produce significant $r$.  This is thus evidence for a delay until
the first $r$-producing stars, and thus points to longer-lived
progenitors (lower mass supernovae or NS-NS binaries).

\item
Finally, the points could continue to cluster around 
$[\rfe]_{\rm avg}$.  This would indicate that the earliest
stars could collectively produce {\em both} the \rpro\ or iron, but
the persistent scatter would demand that {\em individual} stars
could only produce one of the two.  This points to a picture
of the kind suggested by Wasserburg \& Qian, wherein an early
(and non-uniform)
Pop III production of iron is quickly followed by rapid
production of iron-free \rpro\ ejecta.

\end{itemize}

Thus we see that ultra-metal-poor stars can provide
``smoking gun'' evidence for \rpro\ astrophysics.

\section{Conclusions}
\label{sect:conclude}

We have presented an analytical approach to understanding the
scatter in \rpro-to-iron ratios in metal-poor halo stars.
The observed \rfe\ scatter demands that
the bulk of the \rpro\ and of iron cannot be
produced in the same stars.
We have used a simple stochastic description of the different
populations of nucleosynthesis events which contribute
to \rpro\ and iron abundances in each halo star.  
The random star-to-star variations in nucleosynthetic
``ancestry'' lead to scatter in \rfe.
The models we present are all successful in 
reproducing the scatter in the available data, which 
go down to about $[\feh] = -3$.
We see that the abundance scatter can be understood simply 
in terms of the variance due to counting statistics of the
ancestry of each star.  
We present simple and convenient analytic expressions
for the scatter in this regime.

We model two basic scenarios for \rpro\ production;
one in which the \rpro\ is rare, occurring in about
$\sim 4\%$ of all massive stars.  The other scenario,
that of WQ2000, takes the reverse approach in which
\rpro\ production is  ubiquitous and iron production
rare in massive stars.\footnote{In the final stages of completion
of this paper, we became aware of the work of Qian \cite{qian2001},
which also considers \rfe\ scatter in the WQ2000 model.
His methods and conclusions are very similar to ours.}
We find that these scenarios give very similar
\rfe\ scatter at $[\feh] \ga -2.5$, but show divergent 
behavior at lower metallicity.
Observations in this regime can discriminate
among these scenarios and give important clues
regarding the lifetimes and thus the masses of
the astrophysical \rpro\ production sites.  
We therefore strongly urge an observational effort to
measure \rpro\ abundances (and upper limits!) in 
the ultra-metal-poor halo stars.

\acknowledgments
We thank Roberto Gallino, 
John Lattanzio, Chris Sneden, and Ron Webbink for useful discussions.
This research has received support from NSF grant
AST-9986974 to J.J.C.
and
from DOE contract B341495 to J.W.T.

\clearpage



\end{document}